\documentclass{article}
\usepackage[english]{babel}
\usepackage{a4}
\newcommand{\abs}[1]{\left|#1\right|}
\newcommand{\be}{\begin{equation}}
\newcommand{\ee}{\end{equation}}

\begin{document}
\title{\bf A proposed modification of the dinamic subgrid scale closure
  method} 
\author{\small{A. Costa$^*$}\\
\small{Istituto Nazionale di Geofisica e Vulcanologia, Napoli, Italy. }}
\date{\small{Napoli, 24-2-2004}.\\
\small{$^*$Now at Environmental Systems Science Centre, University of Reading, UK.}}
\maketitle
A popular choice for modelling turbulence subgrid tensor $\tau_{ij}$
consists of introducing the concept of ``eddy viscosity'' 
$\nu_{turb}$:
\be
\tau_{ij}-\frac{1}{3}\delta_{ij}\tau_{kk}=-2\nu_{turb}\overline{S}_{ij}
\label{eddy}
\ee
where $\overline{S}_{ij}$ is the shear stress filtered on the scale
$\Delta$.  The quantity $\nu_{turb}$ characterizes the turbulence. Its
order of  magnitude is given by the product of a length times a velocity 
characteristic of the turbulent motion: $\nu_{turb}\sim l \cdot V_l $.
Choosing the length equal to the computational grid, $l=\Delta$, and
$V_\Delta=\abs{\overline{S}}\Delta$, we obtain the Smagorinsky model 
\cite{sma63}: 
\be 
\nu_{turb}=C \Delta^2 \abs{\overline{S}}
\label{tghos}
\ee
The quantity $C$ can be calculated dynamically from
the Germano identity:
\be
\tau_{ij}^{2\Delta}-\widehat{\tau_{ij}^{\Delta}}=L_{ij}
\label{germano}
\ee
which furnishes:
\be
-2\Delta^{2}C=\frac{\langle L_{ij}\widehat{\overline{S_{ij}}}\rangle}
{\langle\widehat{\overline{S_{ij}}}\widehat{\overline{S_{ij}}}\rangle}  
\label{ckol}
\ee
The behaviour of the parameterization (\ref{ckol}) is very similar to
the dynamic Smagorinsky model formulated in \cite{lil92}. Hence the
tensor given by equation (\ref{ckol}) is:  
\be
\tau_{ij}-\frac{1}{3}\delta_{ij}\tau_{kk} =
\frac{\langle L_{ij}\widehat{\overline{S_{ij}}}\rangle} 
{\langle\widehat{\overline{S_{ij}}}\widehat{\overline{S_{ij}}}\rangle}  
{\overline{S}}_{ij}     
\label{taunos1}
\ee
In order to eliminate the constrain that $\tau_{ij}$ is parallel to
$S_{ij}$, we propose the following parameterization:
\be
\tau_{ij}=-2\Delta^{4/3} (\overline{S} C_{ij}+c \overline{S}_{ij})
\label{taunos}
\ee 
where $\overline{S}$ is a quantity representative of the strain-rate 
(for instance one invariant of the tensor $\overline{S}_{ij}$). The
tensor components $C_{ij}$ are calculated from Germano's
identity directly, without introducing the concept of eddy viscosity,
wheras the scalar quantity $c$ is the one that better
reproduce the dissipation energy.\\
Combining (\ref{taunos}) and (\ref{germano}) we obtain:  
\be
-2(2\Delta)^{4/3}(\widehat{\overline{S}}C^{(2\Delta)}_{ij}+
c\widehat{\overline{S}}_{ij})
+2\Delta^{4/3}(\widehat{\overline{S}C_{ij}} +
c\widehat{\overline{S}}_{ij}) = L_{ij}
\label{l3}
\ee
Assuming that $C^{(2\Delta)}_{ij}=C^{(\Delta)}_{ij}$ and
$\widehat{\overline{S}C_{ij}}\simeq\widehat{\overline{S}}C_{ij}$, we 
have:     
\be
-2\Delta^{4/3}(\widehat{\overline{S}}C^{(2\Delta)}_{ij}+
c\widehat{\overline{S}}_{ij}) = a L_{ij}
\label{l4}
\ee
\be
-2\Delta^{4/3} C_{ij}=a\frac{L_{ij}}{\widehat{\overline{S}}}+
2\Delta^{4/3} c\frac{\widehat{\overline{S}}_{ij}} 
{\widehat{\overline{S}}} 
\label{l5}
\ee
In order to calculate $c$ we multiply eq.(\ref{l4}) by
$\widehat{\overline{S}}_{ij}$ and assume that the following
relationship is valid for the average quantities:  
\be
\langle\widehat{\overline{S}} C_{ij}\widehat{\overline{S}}_{ij}
\rangle \simeq c \langle
\widehat{\overline{S}}_{ij}\widehat{\overline{S}}_{ij} \rangle
\label{aver}
\ee
Using the above relationship we have:
\be
2\Delta^{4/3} c =\frac{a}{2} 
\frac{\langle L_{ij}\widehat{\overline{S}}_{ij} \rangle} 
{\langle \widehat{\overline{S}}_{ij}\widehat{\overline{S}}_{ij} \rangle}
\ee
Hence the tensor  $\tau_{ij}$ becomes:
\be
\tau_{ij}=a\frac{\overline{S}}{\widehat{\overline{S}}}L_{ij} +
\frac{a}{2} \frac{\langle L_{ij}\widehat{\overline{S}}_{ij} \rangle} 
{\langle \widehat{\overline{S}}_{ij} \widehat{\overline{S}}_{ij}
  \rangle}\left(\overline{S}_{ij}-
\frac{\overline{S}}{\widehat{\overline{S}}} 
\widehat{\overline{S}}_{ij}\right)
\ee
and assuming $\overline{S}=\sqrt{\overline{S}_{ij}\overline{S}_{ij}}$
we obtain the final form:
\be
\tau_{ij}=a\frac{\sqrt{\overline{S}_{ij}\overline{S}_{ij}}}
{\sqrt{\widehat{\overline{S}}_{ij}\widehat{\overline{S}}_{ij}}}
L_{ij} +
\frac{a}{2} \frac{ \langle L_{ij}\widehat{\overline{S}}_{ij} \rangle} 
{\langle \widehat{\overline{S}}_{ij} \widehat{\overline{S}}_{ij}
  \rangle}\left(\overline{S}_{ij}-
\frac{\sqrt{\overline{S}_{ij}\overline{S}_{ij}}} 
{\sqrt{\widehat{\overline{S}}_{ij}\widehat{\overline{S}}_{ij}}}
\widehat{\overline{S}}_{ij}\right)
\label{tnos}
\ee
We can see that the tensor (\ref{tnos}) contains the first term that
is similar to the tensor predicted by similarity models
\cite{bardina}: 
\be
\tau_{ij}=C_{sim}L_{ij}
\label{tsim}
\ee
and a second term similar to the dynamic Smagorinsky models such as 
(\ref{taunos1}). However, in the case of (\ref{tnos}), the term
proportional to $L_{ij}$ derives from the Germano identity directly.  
It is well-known from the literature, that similarity models better
reproduce the SG tensor components, whereas models based on the concept
of eddy viscosity (\ref{eddy}), such as dynamic Smagorinsky model,
better reproduce dissipated energy \cite{menyeu96}. Tensor
(\ref{tnos}) may be able to better reproduce both SG tensor
components and dissipated energy. 
\section*{Acknowledgments}
{\small M.V. Salvetti is acknowledged for her comments and suggestions.} 

\end{document}